\documentclass[reprint,onecolumn,showpacs,aps,pra,amsmath,amssymb,superscriptaddress,floatfix,notitlepage]{revtex4-1}
\usepackage{graphicx} 
\usepackage{tikz}
\usepackage{placeins}
\usepackage{color}
\usepackage[normalem]{ulem}
\newcommand{\Sys}{\mathcal{S}}
\newcommand{\E}{\mathcal{E}}

\newcommand{\ket}[1]{\vert #1 \rangle}
\newcommand{\bra}[1]{\langle #1 \vert}
\newcommand{\op}[2]{\vert#1\rangle\langle#2\vert}
\newcommand{\inner}[2]{\langle #1 \vert #2 \rangle}
\newcommand{\abs}[1]{\vert #1 \vert}

\newcommand{\tr}[2]{\text{tr}_#2\left\{#1\right\}}
\newcommand{\Hil}{\mathcal{H}}

\newcommand{\diff}{{\rm d}}

\begin{document}
\title{Discrete dynamics and non-Markovianity}

\author{Kimmo Luoma}
\email{kimmo.luoma@tu-dresden.de}
\affiliation{Institut f{\"u}r Theoretische Physik, Technische Universit{\"a}t Dresden, 
D-01062,Dresden, Germany}
\author{Jyrki Piilo}
\affiliation{Turku Centre for Quantum Physics, Department of Physics and 
Astronomy, University of Turku, FI-20014, Turun Yliopisto, Finland}




\date{\today}

\begin{abstract}
We study discrete quantum dynamics where single evolution step consists of unitary system transformation followed by  
decoherence via coupling to an environment. Often non-Markovian memory effects are attributed to structured environments 
whereas here we take a 
more general approach within discrete setting. In addition of controlling the structure of the environment, we are interested 
in how local 
unitaries on the open system allow the appearance and control of memory effects.  Our first simple qubit model, where local 
unitary is followed by 
dephasing, illustrates how memory effects arise despite of having no-structure in the environment the system is coupled with.
We then elaborate this observation by constructing a model for open quantum walk where the unitary coin and transfer operation 
is augmented with 
dephasing of the coin. 
The results demonstrate that in the limit of strong dephasing within each evolution step, the combined coin-position open 
system always displays 
memory effects and their quantity is independent of the structure of the environment.
Our construction makes possible an experimentally realizable open quantum walk with photons exhibiting non-Markovian features.

\end{abstract}

\maketitle
\section{Introduction}

In recent years, there has been a  growing interest on defining and quantifying 
quantum memory effects in open system dynamics~\cite{wolf_assessing_2008,breuer_measure_2009,
rivas_entanglement_2010,lu_quantum_2010,luo_quantifying_2012,
lorenzo_geometrical_2013,bylicka_non-markovianity_2014,chruscinski_degree_2014,
buscemi_equivalence_2014,breuer_non-markovian_2015}. 
These are based on a number of different 
approaches, ranging, e.g., from the concepts of information 
flow~\cite{breuer_measure_2009}, to non-divisibility~\cite{rivas_entanglement_2010,chruscinski_degree_2014}  
and mutual information~\cite{luo_quantifying_2012}. Based on these theoretical developments, experimental realizations for 
detecting and 
controlling of non-Markovianity has become feasible \cite{liu_locality_2014,liu_experimental_2011} and also a number of proposals to exploit memory 
effects, e.g., to quantum 
information tasks has been recently proposed  \cite{vasile_continuous-variable_2011,bylicka_non-markovianity_2014,bylicka_thermodynamic_2015}. 
The previous studies on non-Markovian quantum dynamics mostly focus on continuous coupling between the open system and its 
environment. However, 
there also exists other possibilities to study memory effects, for example discrete dynamics which we consider here. 
In this case, the open system couples in stepwise manner in time to its environment. One can also envisage a possibility of 
having unitary 
transformations changing the state of the open systems between non-unitary evolution steps caused by the environment. 
We are interested in what is the state of the open system after each step which consists consecutive unitary and non-unitary 
parts. 
Thereby, the theoretical challenge is to construct a discrete dynamical map for the stepwise evolution of the system of 
interest and this map 
should contain information about the local unitary within the system of interest and the properties of the environment it is 
interacting with.  

Quantum walks provide a promising base to combine the study of discrete dynamics and memory effects.
Generally, in quantum walks the system can evolve discretely or 
continuously \cite{venegas-andraca_quantum_2012,aharonov_quantum_2000,kempe_quantum_2003}
and a study on the relation between the two cases can be found in 
\cite{strauch_connecting_2006}.
Quantum walks have proven to be important in such diverse fields as quantum information processing 
\cite{lovett_universal_2010}, complex networks \cite{caruso_universally_2014,faccin_degree_2013} and the 
physics of topological phases \cite{kitagawa_topological_2012,puentes_observation_2014}. 
Moreover, the dynamics of quantum walks show a very broad range of different dynamical behavior from ballistic 
spreading to localization \cite{ghosh_simulating_2014,crespi_anderson_2013,chandrashekar_disorder_2012,
jackson_quantum_2012,schreiber_decoherence_2011,ahlbrecht_disordered_2011,ahlbrecht_asymptotic_2011,
kendon_quantum_2006,tregenna_controlling_2003}. Introducing noise to the quantum walk, the effects of 
the transition from unitary to non-unitary dynamics and the classical limit can be studied
\cite{venegas-andraca_quantum_2012,attal_open_2012,attal_open_2012-1,schreiber_decoherence_2011,ahlbrecht_asymptotic_2011,
lavicka_quantum_2011,fan_convergence_2011,broome_discrete_2010,maloyer_decoherence_2007,kendon_decoherence_2007,
kosik_quantum_2006,
ermann_decoherence_2006,brun_quantum_2003,brun_quantum_2003-1,brun_quantum_2003-2,kendon_decoherence_2003,
kendon_decoherence_2002}. 
Recently, quantum walks have been implemented 
experimentally for example using trapped ions \cite{zahringer_realization_2010,karski_quantum_2009}, 
atoms in optical lattices \cite{robens_ideal_2015},
linear optics \cite{puentes_observation_2014,broome_discrete_2010}, optical fibers \cite{schreiber_decoherence_2011,
schreiber_photons_2010} and 
waveguides \cite{crespi_anderson_2013,sansoni_two-particle_2012,perets_realization_2008}. 

In this article we study non-Markovian discrete dynamics of discrete quantum walk in a line.
For this purpose, we introduce a new type open quantum walk where the dynamics is characterized by 
a non-divisible discrete dynamical map. 
It is worth pointing out that, to the best of our knowledge, 
all previous approaches to decoherent quantum walks can be characterized by using discrete quantum dynamical
semigroup. We go beyond dynamics which is describable by semigroup and are interested in how to induce and control memory 
effects for quantum 
walks. The article is structured in the following way.
We first introduce the concept of discrete dynamical map and also formulate the suitable quantifier for memory effects. 
To understand better the origin of memory effects in the considered models, we 
then review the standard dephasing model for a qubit with non-Markovian dynamics. We then study discrete open qubit dynamics 
with unitary control 
operation and show how the addition of the control transforms the dynamics of the open system inherently non-Markovian.
The elaborate on the insights obtained from the simple qubit model, we construct
a discrete quantum walk where we identify the coin operator as the local control operation and proceed 
with memory effects in 1-d discrete open quantum walk. Throughout this paper the interaction between 
the open system and its environment is of pure dephasing type and in the quantum walk model the coin is coupled to an environment.
The formulation follows the path and emphasis the experimental realization of the models with photons.


\section{Discrete dynamical map}\label{sec:discrete-dynamics}
Quantum dynamics is generated by quantum dynamical
maps, a family of completely positive and trace preserving
(CPT) maps $\Phi_ n$ , $n\geq 0$, such that $\Phi_0 =\mathbb{I}$ . If the state of
the quantum system, described by a density operator or matrix, is initially $\rho_0$ then $\rho_n=\Phi_n(\rho_0)$ defines
the evolution of the state. Discrete dynamics for a quantum
system emerge if the parameter $n$ indexing the family of
CPT maps $\Phi_n$ takes discrete values only, eg. $n \in \mathbb{N}$.
We assume that $0\in \mathbb{N}$.

CPT maps suitable for studying discrete dynamics 
can be constructed by using the Stinespring's dilation theorem \cite{heinosaari2011mathematical}.
It states that for every CPT map it is possible to assign an unitary 
evolution in enlarged Hilbert space. 
Total Hilbert space is thus $\Hil=\Hil_\Sys\otimes\Hil_\E$ and the map 
is generated as $\rho_n={\rm tr}_{\E}\{U^n \rho_0\otimes\chi(U^\dagger)^n\}$.
Here $U$ is unitary operator acting on total Hilbert space  
and $\chi$ is fixed initial state on the auxiliary Hilbert space. This
construction guarantees that $\Phi_n$ is CPT for each $n$.
Physical interpretation for this construction in the context of 
open quantum systems is that $\Hil_\E$
is a Hilbert space for external environment to which the system is coupled
unitarily.

Quantum dynamical maps can be classified in the following way by their 
divisibility properties.
If the dynamical map $\Phi_n$ satisfies the following decomposition law 
\begin{align}\label{eq:1}
  \Phi_{n+m}(t)=& \Phi_n\circ\Phi_m,
\end{align}
for all $n\in\mathbb{N}$, then the dynamical map is Markovian and forms a 
discrete dynamical semigroup.
The dynamical map is called CP-divisible if 
\begin{align}\label{eq:2}
  \Phi_{n+m}=W_{n+m,m}\circ\Phi_{m},
\end{align}
for $n,m\in\mathbb{N}$ and where $W_{n+m,m}$ is completely positive and trace
preserving.
If the dynamical map can be 
composed as 
\begin{align}
  \Phi_{n+m}=&V_{n,m}\circ\Phi_{m},
\end{align}
where $V_{n,m}$ is positive and trace preserving then the dynamical
map is called P-divisible \cite{vacchini_markovianity_2011}.

It is clear that the divisibility property of the dynamical map goes beyond the concept of semigroup. However, the 
various definitions of quantum non-Markovianity are still under active 
discussion \cite{breuer_non-markovian_2015}.
In the next section we will introduce the measure used 
in this work to
quantify non-Markovianity of the discrete dynamical map.

\section{Measure for non-Markovianity}\label{sec:meas-non-mark}
Following~\cite{breuer_measure_2009},
we use a measure based on distinguishability of quantum states, quantified
by the trace distance 
$d(\rho_1,\rho_2)=\frac{1}{2}{\rm tr}\abs{\rho_1-\rho_2}$.
Trace distance is contractive under positive and trace preserving maps~\cite{perez-garcia_contractivity_2006}, hence
also under completely positive and trace preserving maps.
Thus, the evolution of trace distance for fixed initial pair under 
P-divisible map is contractive, which means that 
$D_{\rho_1,\rho_2}(m)\geq D_{\rho_1,\rho_2}(n)$ for all $m<n$ where 
$D_{\rho_1,\rho_2}(n)=d(\Phi_n\rho_1,\Phi_n\rho_2)$.
If we find that the trace distance increases, $D_{\rho_1,\rho_2}(n)>D_{\rho_1,\rho_2}(m)$ for some 
$n>m$, then we know that the dynamical map is not P-divisible and we say
that the dynamics is non-Markovian. 

To construct a measure for discrete dynamics, we define the increment of the trace 
distance evolution
\begin{align*}
  \Delta_{1,2}(n)=&D_{\rho_1,\rho_2}(n)-D_{\rho_1,\rho_2}(n-1),\, n\geq1,\\
  \Delta_{1,2}(0)=&0,
\end{align*}
and the measure for non-Markovianity
is defined in terms of the increment as
\begin{align}\label{eq:3}
  \mathcal{N}(\Phi)(n)=&
  \max_{\rho_1,\rho_2}\sum_{n\in S=\{n\in \mathbb{N}\vert \Delta_{1,2}(n)>0\}} \Delta_{1,2}(n)
\end{align}
It can be shown that it is sufficient to make the maximization  
over orthogonal pairs of states ~\cite{wismann_optimal_2012} 
and it also has a local representation \cite{liu_locality_2014}.
The measure has physical interpretation in terms of information
flow between the system and the environment. When 
$\Delta_{1,2}(n)<0$ information flows away from the system to the 
environment and the ability to distinguish the two states 
decreases. When $\Delta_{1,2}(n)>0$ there is a backflow of information 
from the environment to the system which improves the distinguishability.
In general, a lower bound for non-Markovianity is 
quite straightforward to obtain using only a small number of initial states.
It is also worth noting that there exists evolutions 
which are not P-divisible but the trace distance 
between evolving states might still decrease at all points of time. 
In this work we do not present the results for full optimization over all the initial states 
but instead focus on specific pairs of initial states which allow
to witness the non-Markovianity of the dynamics.

\section{Discrete qubit dynamics}\label{sec:discr-qubit-dynam}
\subsection{Dephasing}\label{sec:dephasing}
Genuine quantum effect on open quantum system dynamics is the loss of quantum coherences without
energy exchange between the system and the environment. This effect is called pure dephasing.
Our motivation to study this effect on qubit dynamics is due to possibility of experimental implementation
using optical elements~\cite{liu_experimental_2011}. 
Coupling between the system and the environment is given 
by 
\begin{align}\label{eq:20}
  U_{\delta t} =& \int \diff\omega\sum_{\nu=L,R}e^{in_\nu\omega \delta t}
  \op{\nu}{\nu}\otimes\op{\omega}{\omega},
\end{align}
where $\ket{\nu}$ labels the qubit degree of freedom and $\ket{\omega}$
corresponds to the environmental degree of freedom. With an optical implementation in mind,
this type of unitary dynamics describes the interaction of a photon with a birefringent medium, eg.
quartz \cite{liu_experimental_2011}.  
Different basis states $\ket{\nu}$ correspond to the different polarization states,
$n_\nu$ is the polarization dependent index of refraction, $\omega$ is the frequency of the 
photon and $\delta t$ corresponds to the thickness of the quartz plate.

For a fixed 
product initial state $\varrho=\rho(0)\otimes\chi$ this coupling 
generates pure dephasing dynamics, 
given by a dynamical map  $\Phi_{\delta t}^{\rm PD}\rho(0)=\tr{U_{\delta t}\varrho U_{\delta t}^\dagger}{\E}$, expressed in 
$\ket{\nu}$ basis as
\begin{align}\label{eq:21}
  \Phi_{\delta t}^{\rm PD}:&\left\{\begin{array}{ll}
      \op{\nu}{\nu}&\mapsto\op{\nu}{\nu},\,\forall \nu,\\
      \op{L}{R}&\mapsto\kappa(\delta t)\op{L}{R},\\
      \op{R}{L}&\mapsto\kappa^*(\delta t)\op{R}{L}
      \end{array}
      \right.
\end{align}
The map is determined by the the function $\kappa:\mathbb{R}_+\mapsto\mathbb{C}$,
$\kappa(\delta t)=\int\diff \omega e^{i\Delta n \omega \delta t}\abs{\chi(\omega)}^2$, where 
$\Delta n = n_L-n_R$.
Throughout this work we take the environment initial population distribution $\abs{\chi(\omega)}^2$
(spectrum) to be constructed of two Gaussians with widths $\sigma$, amplitudes $\frac{A}{1+A}$ 
and $\frac{1}{1+A}$, where 
$A\in[0,1]$, central frequencies of the peaks $\mu_1,\mu_2$ and peak separation $\delta\omega=\mu_2-\mu_1$, so that
\begin{align}\label{eq:49}
  \abs{\chi(\omega)}^2=&\frac{1}{1+A}\left(\frac{1}{\sqrt{2\pi\sigma^2}}\right)\left(e^{(\omega-\mu_1)^2/(2\sigma^2)}
                         +Ae^{(\omega-\mu_2)^2/(2\sigma^2)}\right).
\end{align}
This choice is experimentally motivated \cite{liu_experimental_2011} and with parameters 
$A$ and $\sigma$ it is possible to control the structure of the environment.

\begin{figure}
  \includegraphics[scale=1.0]{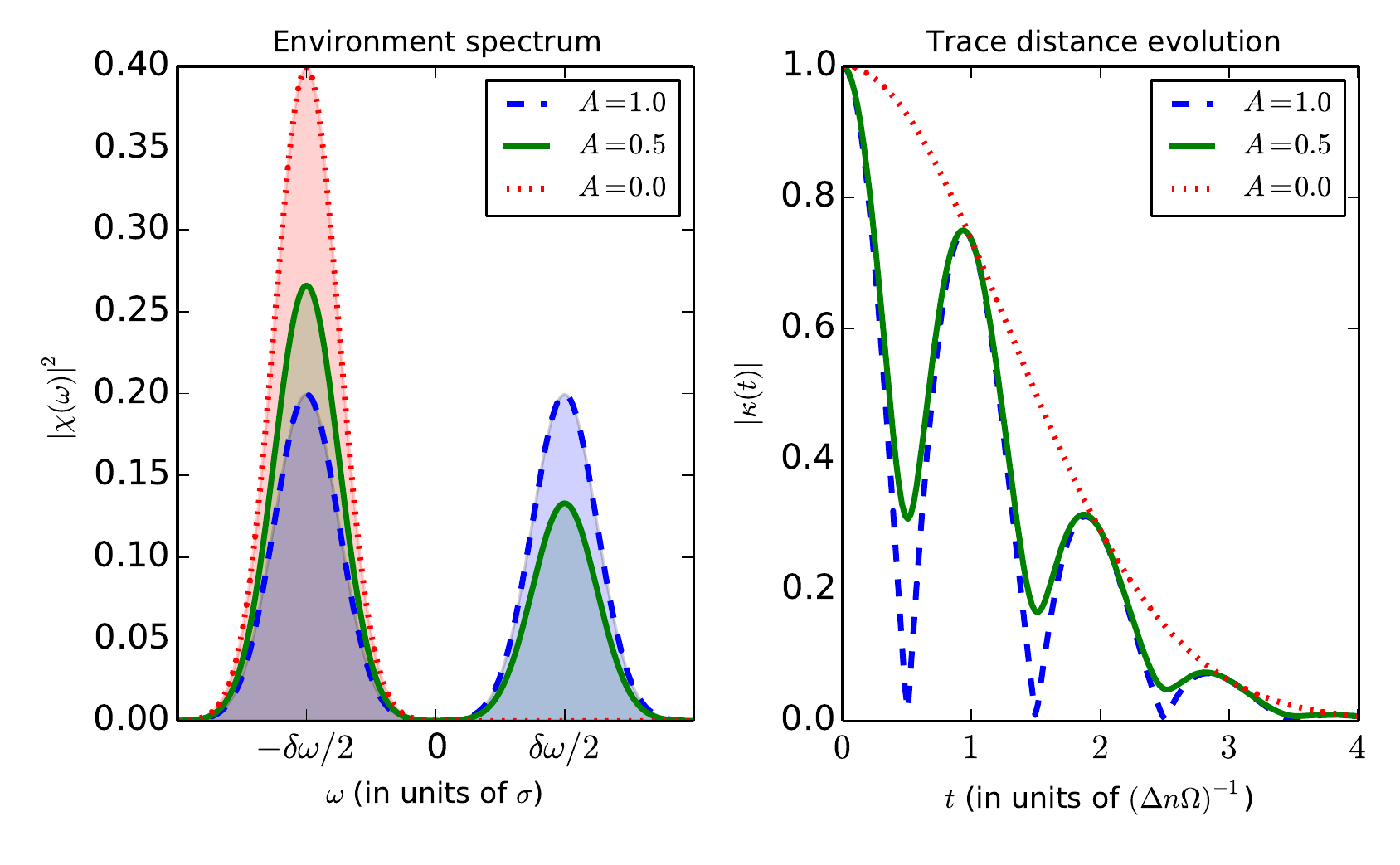}
  \caption{\label{fig:Dephasing} Standard dephasing model for a qubit. On the left is plotted 
  the spectrum of the environment, it consists of Gaussian peaks. 
  On the right is plotted the trace distance dynamics for 
  maximizing initial state pair. As one can see, the single Gaussian peak gives Markovian
  dynamics and more structured environment gives rise to  quantum memory effects. 
  Here $\Omega = \delta\omega/(2\pi)$.}
\end{figure}

It can be shown that the dynamics over a period $\delta t$ can be
Markovian or non-Markovian depending on the properties of the population
distribution $\abs{\chi(\omega)}^2$ of the initial state of 
the environment. Dynamics as a function of $\delta t\equiv t$ is plotted in 
Fig.~\ref{fig:Dephasing}. 
Trace distance dynamics for the optimal pair
of initial states is given by 
$D_{\rho_1,\rho_2}(\delta t)=\abs{\kappa(\delta t)}$. 
This type of dynamics is 
analyzed in \cite{liu_experimental_2011}.

\subsection{Dephasing with local control}\label{sec:dephasing-with-local}

We consider now discrete dynamics where at each step in the dilatation space 
we have an action of a local control unitary operation~\cite{berglund_quantum_2000}
followed by dephasing. The
single step unitary in the total space is then
\begin{align}\label{eq:19}
  V = U_{\delta t} \cdot (C_{\eta}\otimes I_\E).
\end{align} 
In this work we consider only the following local control unitaries $C_{\eta}$
\begin{align}\label{eq:4}
  C_{\eta} =&\sqrt{\eta}(\op{L}{L}-\op{R}{R})+\sqrt{1-\eta}(\op{L}{R}+\op{R}{L}).
\end{align}
These correspond to biased beam splitter transformations and 
$\eta=\frac{1}{2}$ being the Hadamard transformation also known as 
the balanced beam splitter transformation.
The local unitary operator transforms the polarization basis into a new basis
which is not generally simultaneously diagonalizable with the decoherence basis. 
This proves to be crucial for the non-Markovianity of the quantum dynamics as we will show.
The reduced dynamics generated by $V^n$ is given by 
$\Phi_n^{\rm Q}$
which is defined as 
\begin{align}\label{eq:5}
  \rho(n)=\Phi_n^{\rm Q}\rho=&\tr{V^n(\rho\otimes\chi)(V^\dagger)^n}{\E}.
\end{align}

The state of the qubit can be expressed as 
$\rho=\frac{1}{2}(\mathbb{I}_2+\vec{r}\cdot\vec{\sigma})$, where 
$\vec{r}=(r_1,r_2,r_3)^T$ is the Bloch vector and 
$\vec{\sigma}=(\sigma_1,\sigma_2,\sigma_3)^T$ where $\sigma_i$ are the usual Pauli matrices. 
For arbitrary values of $\eta$ the dynamics  is most conveniently given in terms 
of the Bloch vector. The dynamical map can be  written as 
\begin{align}\label{eq:6}
  \vec{r}(m)=&\int d\omega \abs{\chi(\omega)}^2 M(\omega)^m\vec{r}.\\ 
\end{align}  
The matrix $M(\omega)$ is given by 
\begin{align}\label{eq:7}
  M(\omega)=&
  \begin{pmatrix}
    -\beta\cos{\Delta n \delta t \omega } & 
    -\sin(\Delta n \delta t \omega ) & 
    \alpha\cos{\Delta n \delta t \omega } \\
    \beta \sin{\Delta n \delta t \omega } & 
    -\cos{\Delta n \delta t \omega} & 
    -\alpha\sin{\Delta n \delta t \omega} \\
    \alpha & 0 & \beta 
  \end{pmatrix},
\end{align}
where $\alpha = 2\sqrt{(1-\eta)\eta}$, $\beta=2\eta-1$ and 
$\Delta n = n_L-n_R$.
Matrix $M(\omega)$ is periodic, eg. $M(\omega+\tilde\Omega)=M(\omega)$, where
$\tilde\Omega=\frac{2\pi}{\delta t\Delta n}$. 
Numerical integration of 
$M(\omega)^m$ must be done carefully since integrands are highly oscillatory. However, reliable numerical results 
can be obtained. In the strong dephasing limit, $\tilde\Omega\ll \sigma$ and for 
some particular values of $\eta$ it is possible to obtain analytical results.

\begin{figure}
  \includegraphics[scale=1.0]{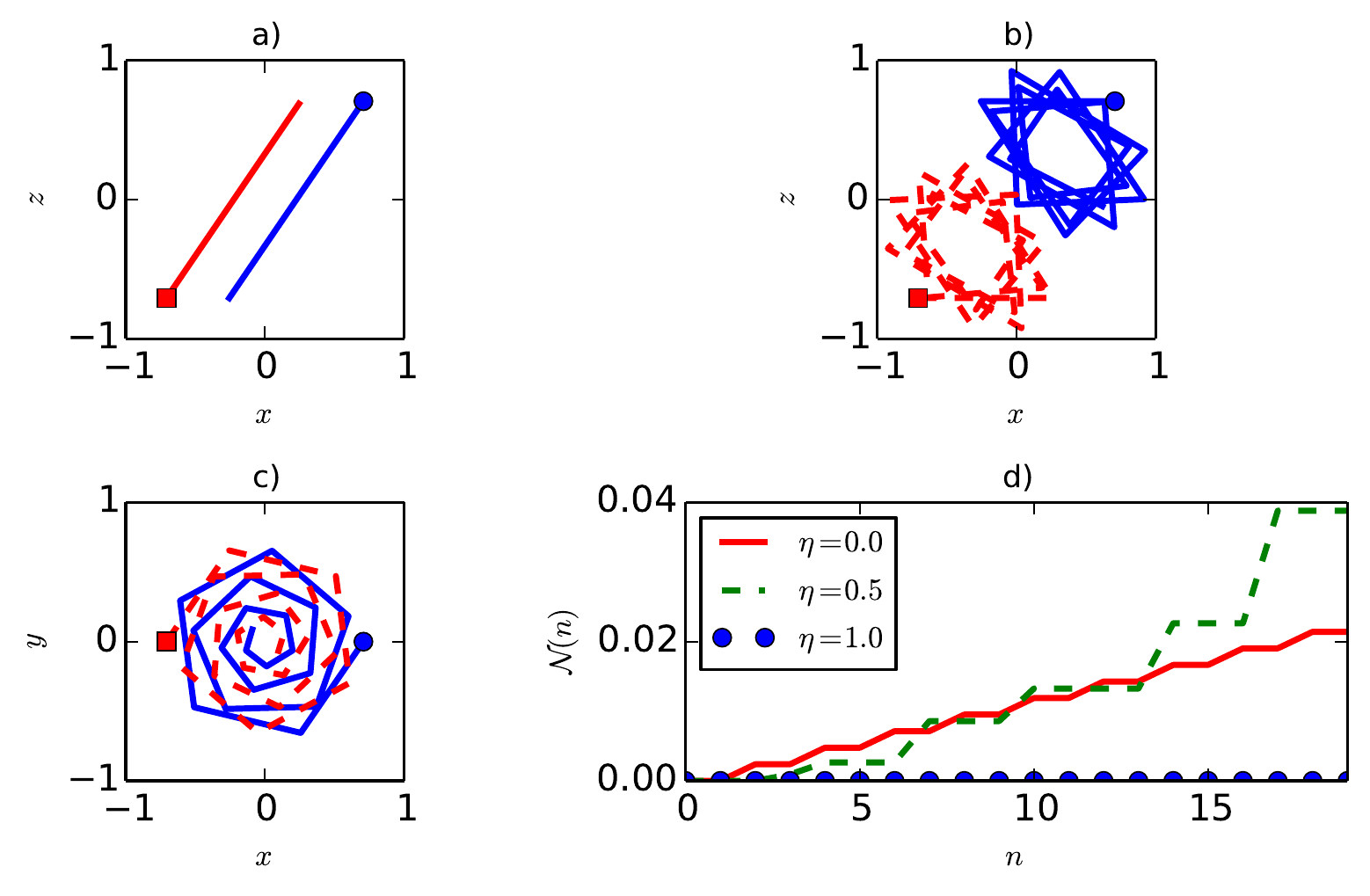}
  \caption{\label{fig:bsqp_weak} Weak dephasing interaction.
  Panel a) $\eta = 0$, b) $\eta = 0.5$, c) $\eta = 1.0$ and we have plotted the time evolution of non-zero Bloch vector components. 
  Panel d) plots the measure for non-Markovianity for fixed pair of initial states for the used three values 
  $\eta$. 
  Initial states in all figures are $\vec{r}_1=\frac{1}{\sqrt{2}}(1,0,1)^T$ and 
  $\vec{r}_2=-\vec{r}_1$ expressed in terms of the Bloch vector. 
  Parameters of the environment are $A=0$ and 
  $\delta \omega = 9\sigma$. Parameters for the interaction are  $\Delta n = 0.009$ and 
  $\delta t = 0.014\frac{2\pi}{\delta \omega \Delta n}$.
  This gives $\frac{\tilde{\Omega}}{\sigma}\approx 643\gg 1$.}
\end{figure}

\begin{figure}
  \includegraphics[scale=1.0]{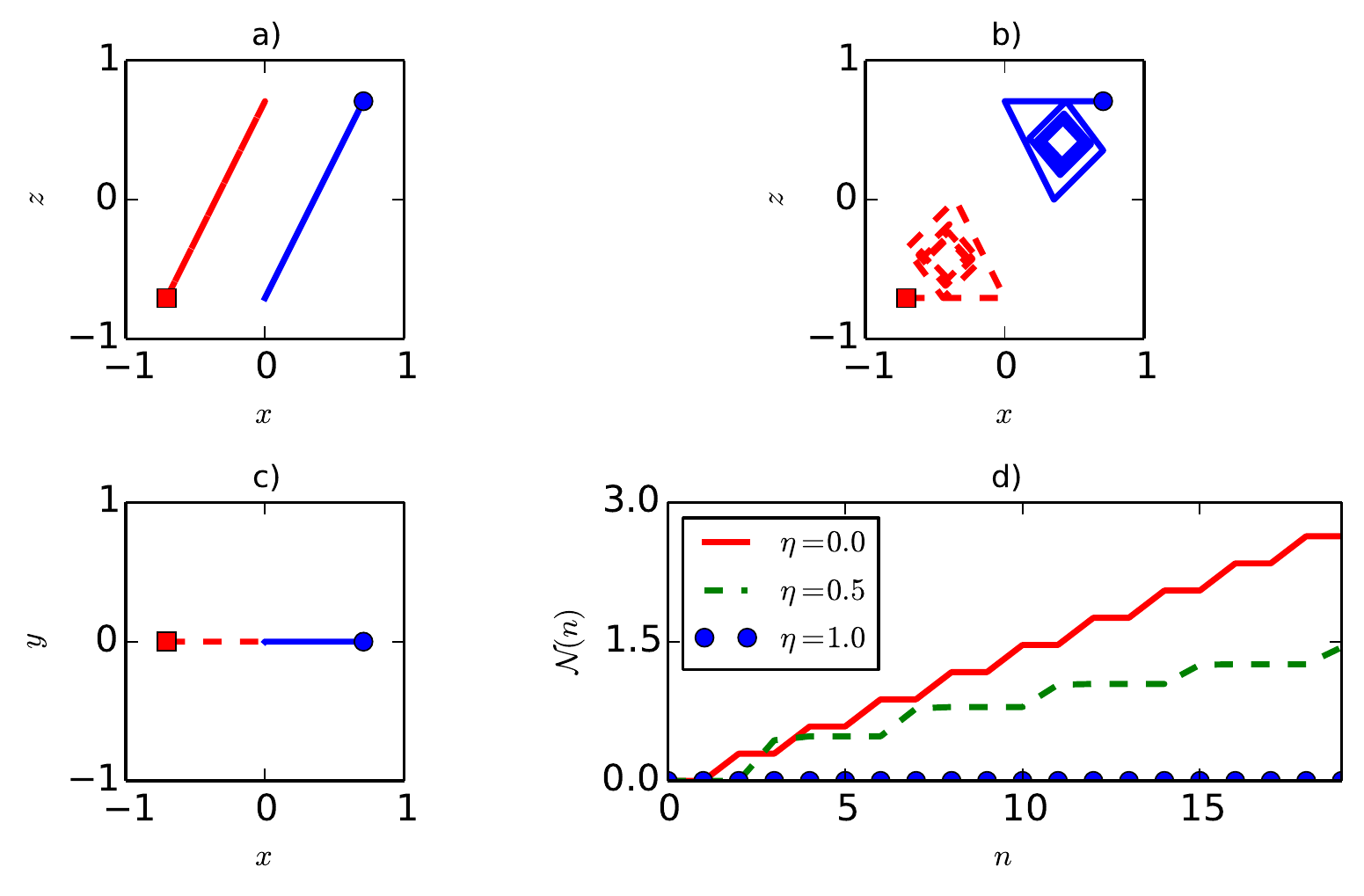}
  \caption{\label{fig:bsqp_strong} Intermediate dephasing interaction.
    Panel a) $\eta = 0$, b) $\eta = 0.5$, c) $\eta = 1.0$ and we have plotted the time evolution of non-zero Bloch vector components. 
  Panel d) plots the measure for non-Markovianity for fixed pair of initial states for the used three values 
  $\eta$. 
    Initial states in all figures are $\vec{r}_1=\frac{1}{\sqrt{2}}(1,0,1)^T$ and 
    $\vec{r}_2=-\vec{r}_1$ expressed in terms of Bloch vector. 
    Parameters of the environment are $A=0$ and
    $\delta \omega = 9\sigma$. Parameters for the interaction are  $\Delta n = 0.009$ and 
    $\delta t = 2\frac{2\pi}{\delta \omega \Delta n}$.
    This gives $\frac{\tilde{\Omega}}{\sigma} = 4.5 \sim 1$.}
\end{figure}

Figures ~\ref{fig:bsqp_weak} and \ref{fig:bsqp_strong} present the results for the values
$\eta=0,0.5,1.0$ in the case of weak and intermediate dephasing strength.  
It is worth noting that for all of the cases we have $A=0$, i.e., the environmental spectrum is flat corresponding to Markovian dynamics {\it without} local control.

For $\eta=1$, the  local control is $C_{1}=\sigma_z$ and
we have 
\begin{align}
  \bra{\nu,\omega}
  (U_{\delta t}\cdot(\sigma_z\otimes\mathbb{I}_\E))^m
  \ket{L,\omega}=&\delta_{\nu,L}e^{i n_L\omega m \delta t},\label{eq:9}\\
  \bra{\nu,\omega}
  (U_{\delta t}\cdot(\sigma_z\otimes\mathbb{I}_\E))^m
  \ket{R,\omega}=&(-1)^m\delta_{\nu,R}e^{i n_R\omega m \delta t}\label{eq:10}.
\end{align}
This leads to the following dynamical map
\begin{align}\label{eq:11}
  \Phi_{m}^{{\rm Q}}:&\left\{\begin{array}{ll}
      \op{\nu}{\nu}&\mapsto\op{\nu}{\nu},\,\forall \nu,\\
      \op{L}{R}&\mapsto\kappa(m \delta t)(-1)^m\op{L}{R},\\
      \op{R}{L}&\mapsto\kappa^*(m \delta t)(-1)^m\op{R}{L}.
      \end{array}
      \right. 
\end{align}
Compared to the uncontrolled case, the sign of the coherences is flipped. This does not affect
the memory effects since the dephasing process moves the states towards the 
$z$-axis  in the $xy$-plane of the Bloch sphere and the sign change 
of the coherences keeps the distance of the state from the $z$-axis constant.
The dynamics is displayed in panels c) and d) of  Fig.~\ref{fig:bsqp_weak} and Fig.~\ref{fig:bsqp_strong}.

For $\eta=0$ the local control is $\sigma_x$ and we have
 \begin{align}
   \bra{\nu,\omega}
   (U_{\delta t}\cdot(\sigma_x\otimes\mathbb{I}_\E))^{2m}
   \ket{\nu',\omega}=&\delta_{\nu,\nu'}e^{i (n_\nu+n_{\nu'})\omega m \delta t},\label{eq:13}\\
   \bra{\nu,\omega}
   (U_{\delta t}\cdot(\sigma_x\otimes\mathbb{I}_\E))^{2m+1}
   \ket{\nu',\omega}=&(1-\delta_{\nu,\nu'})
                          e^{i(mn_{\nu'}+(m+1)n_\nu)\omega\delta t}.\label{eq:14}
 \end{align}
This gives the dynamical map 
\begin{align}
  \Phi_{2m}^{\rm Q}:& \op{\nu}{\nu'}\mapsto \op{\nu}{\nu'},\label{eq:15}\\
  \Phi_{2m+1}^{\rm Q}:&\left\{\begin{array}{ll}
      \op{\nu}{\nu}&\mapsto\op{\nu'}{\nu'},\,\forall \nu\neq\nu',\\
      \op{L}{R}&\mapsto\kappa(\delta t)^*\op{R}{L},\\
      \op{R}{L}&\mapsto\kappa(\delta t)\op{L}{R}.
      \end{array}
      \right.
\end{align}
In this case, the information flow between the system and the environment is maximal in the sense
that the local control is able to  completely eliminate the effect of the environment 
after even number of steps, see panels a) and d) in Fig.~\ref{fig:bsqp_weak} and Fig.~\ref{fig:bsqp_strong}.

\subsection{Strong dephasing limit}\label{sec:strong-deph-limit-qubit}
Let us assume that we have a ``flat'' population distribution for the initial 
environmental state. What we mean by this is that, $\abs{\chi(\omega)}^2$ stays 
almost constant for $\omega \in [\omega'-{\tilde\Omega}/2,\omega'+\tilde{\Omega}/2]$, where 
$\omega'\in [0,\infty)$.

Now, this means that the effects of the environment in this limit are generic in the sense, that the global
structure of $\abs{\chi(\omega)}^2$ does not play a role. This type of environment is usually called 'Markovian'.
It is intuitively clear that in this type of situation it is enough to integrate only over a single 
period of length $\tilde\Omega$ in Eq.~\eqref{eq:6} to a very good approximation.
In terms of equations this means 
\begin{align}\label{eq:8}
  \vec{r}(m)=& \frac{1}{\tilde\Omega}\int_{0}^{\tilde\Omega}d\omega M(\omega)^m \vec{r}.
\end{align}
Physical idea behind this approximation is that there is no distinguished frequency of the environment because 
of the flatness of the spectrum. Then all the frequencies in the interval of length $\tilde\Omega$ ``mix''
the state of the reduced system with equal weights, hence it is sufficient to consider 
only one of these intervals.

\begin{figure}
  \includegraphics[scale=1.0]{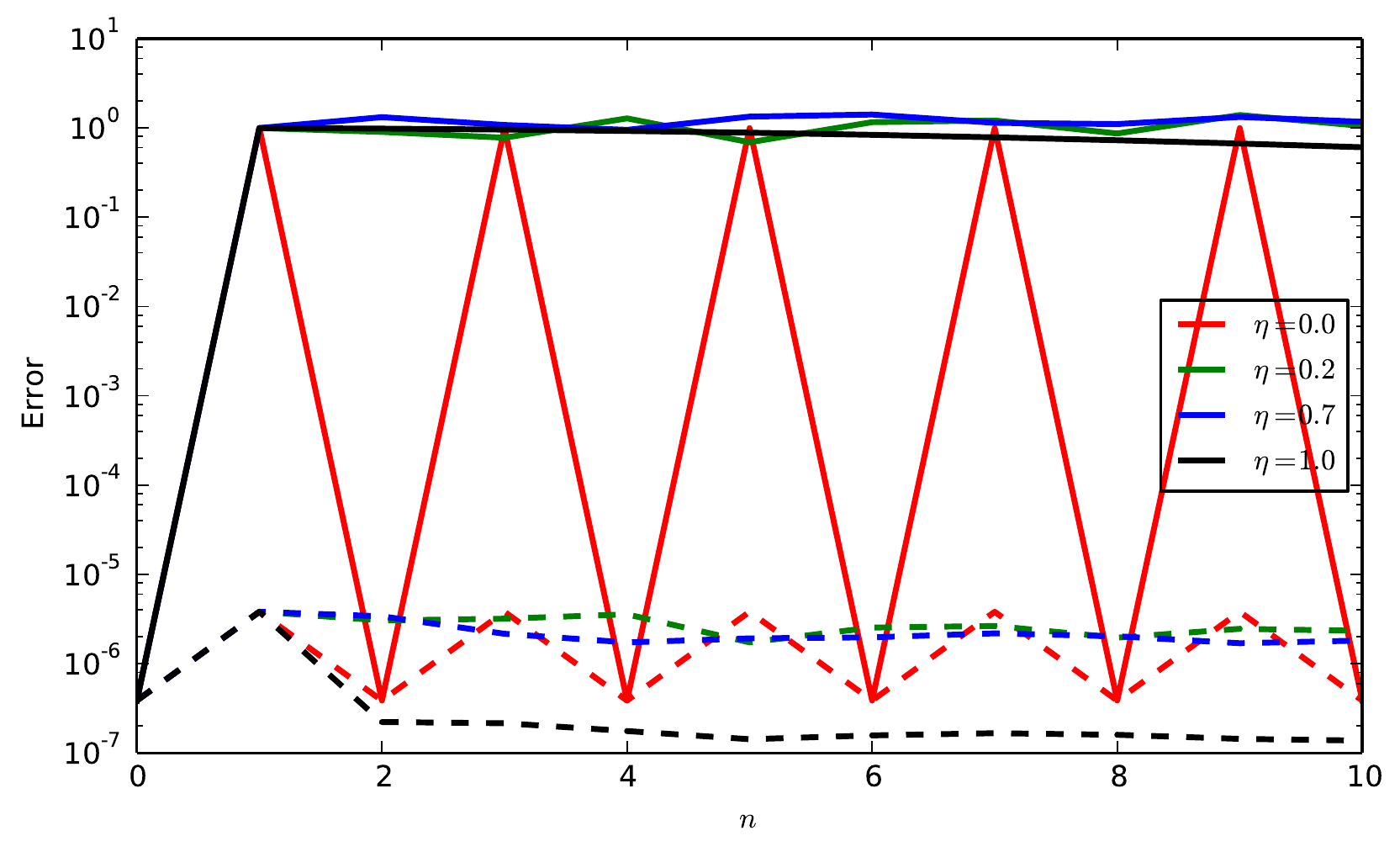}
  \caption{\label{fig:bsqp_error} Error of the strong dephasing approximation for few values 
    $\eta$. Error is measured by $d(\Phi^{\rm Q}_n - \Lambda_n)$, eg. the trace distance between 
    the exact ($\Phi^{\rm Q}_n$) and approximate ($\Lambda_n$) dynamical map. For this calculation 
    we have chosen parameters of the environment as $A=0$ and
    $\delta \omega = 9\sigma$. For interaction we have  $\Delta n = 0.009$ and the 
    solid lines correspond $\delta t=0.02\frac{2\pi}{\delta\omega\Delta n}$ (weak dephasing) 
    and dashed 
    lines correspond to  
    $\delta t=1.03\frac{2\pi}{\delta\omega\Delta n}$ (intermediate dephsing). These corresponds to the 
    following approximate values $\tilde\Omega/\sigma \approx \{450,9\}$ of the quantity measuring the 
    strength of the dephasing.}
\end{figure}

In general we have to validate this approximation numerically, but for the special case 
where $\eta=\frac{1}{2}$ and $A=0$ we can obtain analytical expressions for dynamical 
map in the strong dephasing limit, which we will do next.  It turns out that we will need the condition 
$\sigma\gg\tilde\Omega$ for the analytical calculation but the numerical data shows that 
already for intermediate dephasing, 
$\sigma \sim \tilde\Omega$, strong dephasing approximation~\eqref{eq:8} works quite well, 
see Fig.~\ref{fig:bsqp_error}. Dynamics in the intermediate and strong dephasing 
regime for $\eta=\frac{1}{2}$ is plotted in panel b) of Figs.~\ref{fig:bsqp_weak} and \ref{fig:bsqp_strong}. 

We now give a more detailed proof of Eq.~\eqref{eq:8} for $\eta=\frac{1}{2}$ and $A=0$.
Let $\eta=\frac{1}{2}$ and $\abs{\chi(\omega)}^2 = \frac{1}{\sqrt{2\pi}\sigma}e^{-\frac{(\omega-\mu)^2}{2\sigma^2}}$.
We assume that $\tilde{\Omega}\ll \sigma$, which means that the spectral distribution varies in much larger scale
than the integral kernel $M(\omega)$, which takes the following form
\begin{align}\label{eq:12}
  M(\omega) = 
  \begin{pmatrix}
    0& 
    -\sin(\Delta n \delta t \omega ) & 
    \cos(\Delta n \delta t \omega ) \\
    0& 
    -\cos(\Delta n \delta t \omega) & 
    -\sin(\Delta n \delta t \omega) \\
    1 & 0 & 0 
  \end{pmatrix}.
\end{align}
We decompose the interval $\mathbb{R}_+=[0,\infty)$ as $\bigcup_{k\in \mathbb{N}\setminus\{0\}}\tilde\Omega_k$,
where $\tilde\Omega_k=[\tilde\Omega\cdot(k-1),\tilde\Omega\cdot k)$. 
Since $\int_{\Omega_k}\diff\omega\abs{\chi(\omega)}^2$ is continuous on 
closed interval $\overline{\tilde\Omega_k}$ and differentiable on $\tilde\Omega_k$, mean 
value theorem states that we can always find $\omega_k\in\tilde\Omega_k$ such that
\begin{align}\label{eq:16}
  \abs{\chi(\omega_k)}^2\tilde\Omega=\int_{\tilde\Omega_k}\diff\omega\abs{\chi(\omega)}^2.
\end{align}
As an approximation
we choose the midpoint of each interval,
$\omega_k = \tilde\Omega\cdot k + \frac{\tilde\Omega}{2}$, then
\begin{align}\label{eq:17}
  \sum_{k\in\mathbb{N}\setminus\{0\}}\abs{\chi(\omega_k)}^2\tilde\Omega=&
\vartheta_3(\pi(\frac{1}{2}-\mu/\tilde\Omega),e^{-2\pi^2\sigma^2/\tilde\Omega^2}),
\end{align}
where $\vartheta_3(u,q)$ is Jacobi theta function. 

Condition $\tilde\Omega\ll\sigma$ allows to do the following approximation
\begin{align}\label{eq:18}
  \sum_k\int_{\tilde\Omega_k}\diff\omega 
  \abs{\chi(\omega)}^2 M(\omega)^m
  &\approx \sum_k\abs{\chi(\omega_k)}^2 \int_{\Omega_k}\diff\omega M(\omega)^m
  =\sum_k\abs{\chi(\omega_k)}^2\tilde{M}(m)\notag\\
  &=\frac{1}{\tilde\Omega}\vartheta_3(\pi(\frac{1}{2}-\mu/\tilde\Omega),e^{-2\pi^2\sigma^2/\tilde\Omega^2})\tilde{M}(m),
\end{align}
where $\tilde{M}(m)=\int_{\Omega_k}\diff\omega M(\omega)^m$ and we used Eq.~\eqref{eq:17} in the last step. 
When we take the strong dephasing limit, $\sigma\to \infty$ and use the property $\lim_{q\to 0}\vartheta_3(u,q)=1$ 
of the Jacobi theta function we obtain Eq.~\eqref{eq:8} \cite{abramowitz_handbook_1964}.
Next we will construct the dynamical map in this special case.

\subsection{Dynamical map in strong dephasing limit for $A=0$ and $\eta=\frac{1}{2}$}\label{sec:example-4}
In the strong dephasing limit for single Gaussian spectral distribution and $\eta=\frac{1}{2}$
we obtain the following analytical form for the dynamical map 
$\Lambda_m=\frac{1}{\tilde\Omega}\int_0^{\tilde\Omega} M(\omega)^m$
\begin{align}
  \Lambda_0=&\mathbb{I}, &
  \Lambda_1=&\begin{pmatrix}
    0&0&0\\
    0&0&0\\
    1&0&0
  \end{pmatrix}, \notag\\
  \Lambda_2=&\frac{1}{2}\begin{pmatrix}
    0&0&1\\
    0&1&0\\
    0&0&0
  \end{pmatrix}, &
  \Lambda_3=&\frac{1}{2}\begin{pmatrix}
    1&0&1\\
    0&1&0\\
    0&0&1
  \end{pmatrix},\notag \\
  \Lambda_{2m} =& \begin{pmatrix}
    a_{m-2}&0&a_{m-1}\\
    0&b_{m-1}&0\\
    a_{m-2}&0&a_{m-2}
  \end{pmatrix},m\geq 2, &
  \Lambda_{2m-1}=&\begin{pmatrix}
    a_{m-2}&0&a_{m-2}\\
    0&b_{m-2}&0\\
    a_{m-3}&0&a_{m-2}
  \end{pmatrix},m\geq 3.\label{eq:23}
\end{align}
where 
\begin{align}
  a_k =& \sum_{i=0}^k (2i+1)\frac{C(i)}{(-8)^i}, &
  b_k =& \sum_{i=0}^k \frac{C(i)}{(-8)^i}, \\
  a_k=&b_k=0,\,k<0, &
  C(k) =& \frac{1}{k+1}\binom{2k}{k},\,k\in\mathbb{N}.\label{eq:24}
\end{align}
$C(k)$ is called a Catalan number. $a_k$ and $b_k$ have 
the following limiting behavior
\begin{align}\label{eq:25}
  \lim_{k\to\infty} b_k =& (\sqrt{2}-1), &
  \lim_{k\to\infty} a_k =& 1-\frac{1}{\sqrt{2}}.
\end{align}
Thus the dynamical map takes the following form in the limit of infinite number of steps
\begin{align}\label{eq:26}
  \Lambda_{m\to\infty}=&\begin{pmatrix}
    1-\frac{1}{\sqrt{2}}&0&1-\frac{1}{\sqrt{2}}\\
    0&\sqrt{2}-1&0\\
    1-\frac{1}{\sqrt{2}}&0&1-\frac{1}{\sqrt{2}}
  \end{pmatrix}.
\end{align}

\subsection{Discussion}
The results above show that in a discrete dephasing model for a qubit, the addition of local unitary 
can induce non-Markovian dynamics even for flat "Markovian'' spectral structure.
Local unitary operation can be seen as a periodic control, that dynamically decouples the open 
system from the environment giving rise to a partial revivals of the populations and coherences in the 
open system dynamics. Periodicity of this control operation can be seen from the discreteness 
of the dynamics, i.e., we "watch" the system only at the integer multiples of the 
control period.  It is also worth noting that the local unitary changes the open system state while the earlier created correlations between the 
system and environment still persist. The local change in the system state allows then the existing system-environment correlations to be converted back to 
the increased distinguishability of the system states and backflow of information.

Another effect of the local control, in the case that it does not commute with the 
dephasing basis, e.g. when $C_0=\sigma_x$, is that it transforms the open system dynamics from pure dephasing 
to dissipative. For the case when local control is $C_1=\sigma_z$, which commutes with the dephasing operator $U_{\delta t}$, 
the dynamics is pure dephasing type and the action of the 
local control does not have effect on the non-Markovianity of the dynamics, i.e., non-Markovian dynamics 
can emerge from the spectral structure only.
We also show that the non-Markovianity induced by the local control is generic in the sense that 
the structure of the environment spectrum does not play a role in the strong dephasing 
regime. In the special case of Hadamard control $\eta=0.5$ and $A=0$ we were able to 
derive analytical expression for the dynamical map.

In the following section we study more complicated situation with one dimensional discrete 
quantum walk. Our findings will be better understood with the help of the physical intuition 
gained from the present section.

\section{Open quantum walk}\label{sec:open-quantum-walk}
\subsection{Quantum walk}
Quantum walks are either continuous or discrete time unitary protocols that evolve 
a quantum state on a Hilbert space that is constructed from the underlying 
graph where the walk takes place. In this section we will mostly adapt to the 
notation of Ref.~\cite{reitzner_quantum_2011}.

In this work we limit the discussion to a discrete quantum walks on a line. 
Hilbert space for the walk is 
$\Hil_W = \Hil_C\otimes\Hil_P=\mathbb{C}^2\otimes\ell^2(\mathbb{Z})$. 
It consists 
of the coin- and the position space.
Unitary operator $W$ evolving the state of the walker  
over a single step is the following
\begin{align}\label{eq:32}
  W=& (\op{L}{L}\otimes S +\op{R}{R}\otimes S^\dagger)(C_H\otimes\mathbb{I}_P),
\end{align}
where $S=\sum_x \op{x-1}{x}$ and $C_H=C_{\frac{1}{2}}$, see Eq.~\eqref{eq:4}.
We choose to focus only on
Hadamard walks, meaning that the coin unitary is the Hadamard 
matrix.
The unitary operator $W$ can be diagonalized if we move into a quasi-momentum 
picture by Fourier transform $\ket{k}=\sum_{x=-\infty}^\infty e^{ikx}\ket{x}$,
$k\in [-\pi,\pi)$. $k$ is called quasi-momentum since it is periodic.
Note that the Fourier- or quasi-momentum basis is not normalizable, but nevertheless 
very useful when used carefully. The inverse transformation is defined as 
$\ket{y}=\frac{1}{2\pi}\int_{-\pi}^\pi \diff k\, e^{-iky}\ket{k}$.

In this work we always initialize the position of the walker to the origin.
In the quasi-momentum picture the unitary operator $W$ acts on an 
arbitrary state initialized from origin, $\ket{\phi_0}=\ket{\phi}\otimes\ket{0}$,  as
\begin{align}\label{eq:40}
  \ket{\phi_m}=&W^m\ket{\phi_0}=\int_{-\pi}^\pi\frac{\diff k}{2\pi}\, (M_k)^m\ket{\phi}\otimes\ket{k},
\end{align}
where $M_k$ is the following $2\times 2$ matrix
\begin{align}\label{eq:41}
  M_k=&\frac{1}{\sqrt{2}}\begin{pmatrix} e^{-ik}&e^{-ik}\\e^{ik}&-e^{-ik}
  \end{pmatrix}.
\end{align}
Eigenvalues of $M_k$ are $\{e^{-i\nu_k},-e^{i\nu_k}\}$, where $\nu_k$ is defined by
\begin{align}\label{eq:42}
  \sin k=& \sqrt{2}\sin\nu_k.
\end{align}
Using the quasi-momentum  representation for solving the dynamics and then transforming
back to the position representation, we obtain the following expression for 
a general initial state starting from the origin, $\ket{\psi_0}=(c_L\ket{L}+c_R\ket{R})\otimes\ket{0}$, 
and evolving $m$ steps    
\begin{align}\label{eq:28}
  \ket{\psi_m}=W^m\ket{\psi_0}=&\sum_{x=-m}^m \bigg[\big(c_L A_L^m(x)+c_R A_R^m(x)\big)\ket{L}
                                 +\big(c_L B_L^m(x)+c_R B_R^m(x)\big)\ket{R}\bigg]\otimes\ket{x}.
\end{align}
Analytical expressions for the coefficient functions $A_L^m(x),\,A_R^m(x),B_L^m(x),\,B_R^m(x)$, are obtained 
most easily from the quasi-momentum picture. They are
\begin{align}
  A_L^m(x)=&\frac{1+(-1)^{m+x}}{2}[\alpha^m(x)+\beta^m(x)],&
  A_R^m(x)=&\frac{1+(-1)^{m+x}}{2}[\beta^m(x)-\gamma^m(x)],\label{eq:34}\\
  B_L^m(x)=&\frac{1+(-1)^{m+x}}{2}[\beta^m(x)+\gamma^m(x)],&
  B_R^m(x)=&\frac{1+(-1)^{m+x}}{2}[\alpha^m(x)+\beta^m(x)]\label{eq:36},
\end{align}
where 
\begin{align}
  \alpha^m(x)=&\int_{-\pi}^\pi \frac{\diff k}{2\pi}\,e^{i(kx-m\nu_k)},\label{eq:37}\\
  \beta^m(x)=&\int_{-\pi}^\pi \frac{\diff k}{2\pi} \,\frac{\cos k}{\sqrt{1+\cos²k}}e^{i(kx-m\nu_k)},\label{eq:38}\\
  \gamma^m(x)=&\int_{-\pi}^\pi \frac{\diff k}{2\pi}\,\frac{\sin k}{\sqrt{1+\cos²k}}e^{i(kx-m\nu_k)}\label{eq:39}.
\end{align}
Using the method of stationary phase, asymptotic expressions for Eqs.~\eqref{eq:34} and~\eqref{eq:36}
may be obtained. 

After $m$ number of steps the walker has non-zero probability to be found  in positions $-m,-m+2,\cdots,m-2,m$. From this follows that 
after even number of steps the walker has non-zero probability only at even vertices, similarly for odd number of steps. This property is
sometimes called {\it modularity}.

\subsection{Open quantum walk}\label{sec:open-quantum-walk-1}

We extend the the Hilbert space to take into account the effect of an 
external environment. The extended Hilbert space is 
$\Hil=\Hil_C\otimes\Hil_P\otimes\Hil_\E\equiv \Hil_\Sys\otimes\Hil_\E$ and now 
our system consists of the coin and position degrees of freedom.

We couple the coin degrees of freedom to the external environment and the coupling unitary is given by the 
trivial extension of Eq.~\eqref{eq:20} 
\begin{align}\label{eq:43}
  U_{\delta t} =& \int \diff\omega\sum_{\sigma=L,R}e^{in_\sigma\omega \delta t}\op{\sigma}{\sigma}\otimes\mathbb{I}_P\otimes\op{\omega}{\omega}.
\end{align}
For simplicity we consider only homogeneous case, meaning that the coupling operator acts 
identically on each vertex of the graph. The walk operator is extended to act on the enlarged 
Hilbert space trivially, $W\equiv W\otimes\mathbb{I}_\E$.

The dynamical map for the open quantum walk is constructed as usual
\begin{align}\label{eq:44}
  \Phi_m^W(\rho_0)=& \tr{(U_{\delta t}\cdot W)^m\rho_0\otimes\chi((U_{\delta t}\cdot W)^\dagger)^m}{\E},
\end{align}
where $\chi$ is again initial state for the environment. Remarkably, an analytical expression for the dynamical map can be found.

\subsubsection{Quantum dynamical map}
Matrix elements of the dynamical map are $[\Phi_m^Q]_{\sigma'\sigma,x'x;\tau\tau',yy'}\equiv
\bra{\sigma',x'}\Phi_m^Q(\op{\sigma,x}{\tau,y})\ket{\tau',y'}$. We use the following 
shorthand notation  $[\Phi_m^Q]_{\sigma'\sigma,x';\tau\tau',y'}\equiv[\Phi_m^Q]_{\sigma'\sigma,x'0;\tau\tau',0y'}$.
Matrix elements can be written explicitly by using the definition of the quantum walk
\begin{align}\label{eq:45}
  [\Phi_n^Q]_{\sigma'\sigma,x';\tau\tau',y'}=&\int \diff\omega\,\abs{\chi(\omega)}^2\sum_{\substack{k_n,\cdots,k_1=\{L,R\}\\l_n,\cdots ,l_1=\{L,R\}}}
e^{\left[i\omega\delta t \sum_{s,v=1}^n (n_{k_s}-n_{l_v})\right]}\notag\\
&\times \left(c_{k_n,k_{n-1}}\cdots c_{k_1,\sigma}\right)\left(\bar{c}_{l_n,l_{n-1}}\cdots\bar{c}_{l_1,\tau}\right)
\inner{\sigma'}{k_n}\inner{l_n}{\tau'}\notag\\
&\bra{x'}K_{k_n}\cdots K_{k_1}\ket{0}\bra{0}K_{l_1}^\dagger\cdots K_{l_n}^\dagger\ket{y'},
\end{align}
where $K_L = S$, $K_R=S^\dagger$ and $c_{\sigma,\sigma'}=\bra{\sigma}C_H\ket{\sigma'}$.

Each non-zero inner product $\bra{x'}K_{k_n}\cdots K_{k_1}\ket{0}$ in Eq.~\eqref{eq:45}
corresponds to a path $0\to x'$ in a $n$-level binary tree. Each path $0\to x$ has 
$N_L^n(x)=\frac{n-x}{2}$ and $N_R^n(x)=\frac{n+x}{2}$ left and right turns.
This allows to write the exponential part of Eq.~\eqref{eq:45} as 
\begin{align}\label{eq:46}
  e^{\left[i\omega\delta t \sum_{s,v=1}^n (n_{k_s}-n_{l_v})\right]}=& e^{i\omega\delta t \Delta n (y'-x')},
\end{align}
where $\Delta n = n_L-n_R$, since left and right turns each contribute to the sum  in total 
$N_L(x)$ and $N_R(n)$ times the term $n_L$ and $n_R$, respectively.
Then by suppressing the explicit matrix products we
can write the matrix element as 
\begin{align}\label{eq:47}
  [\Phi_n^Q]_{\sigma'\sigma,x';\tau\tau',y'}=&\int \diff\omega\,\abs{\chi(\omega)}^2
  e^{i\omega\delta t \Delta n (y'-x')}\bra{\sigma',x'}W^n\op{\sigma,0}{\tau,0}(W^\dagger)^n\ket{\tau',y'}.
\end{align}
This shows, that the coupling of the coin degrees of freedom to the external environment 
has an effect only  on the coherences between different sites. It does not have any effect on the 
position distribution or to the propagation speed. 

\subsubsection{Strong dephasing limit}\label{sec:strong-deph-limit-walk}
In the strong dephasing limit the state of the quantum system is 
block diagonal since the coherences between any two different sites in the 
lattice are destroyed. This means that the state of the quantum walker 
can be written as 
\begin{align}\label{eq:48}
  \rho_m=&\bigoplus_{x\in[-m,m]} \tilde{\rho}^m(x),
\end{align}
where $\tilde{\rho}^m(x)=\sum_{\sigma,\sigma'=\{L,R\}}\tilde{\rho}^m_{\sigma,\sigma'}(x)\op{\sigma,x}{\sigma',x}$.
Eigenvalues of a self-adjoint $2\times 2$ matrix $\begin{pmatrix}
a & b\\b^* & c
\end{pmatrix}$ are $\lambda_{\pm}=\frac{1}{2}\left(a+b \pm\sqrt{(a-c)^2+4\abs{b}^2}\right)$.

Pure initial state $\ket{\psi_0}=c_L\ket{L,0}+c_R\ket{R,0}$  evolves in $m$-steps in the strong 
dephasing limit into $\Phi_m^W(\op{\psi_0}{\psi_0})$. Expression for block $x$ is 
\begin{align}
  \bra{x}\Phi_m^W(\op{\psi_0}{\psi_0})\ket{x}=\bra{x}W^m\op{\psi_0}{\psi_0}(W^\dagger)^m\ket{x}\equiv \tilde{\rho}_{\psi_0}(x).
\end{align}
Characterization of non-Markovianity becomes now more feasible, since we need to diagonalize 
at the $m$th step $2m+1$ $2\times 2$ self-adjoint matrices, instead of $m^2$ valued matrix.
Analytical expression by using Eqs.~\eqref{eq:28},\eqref{eq:34},\eqref{eq:36}
could be obtained, but in this work it we do not analyze it further.

\begin{figure}
  \includegraphics[scale=1.0]{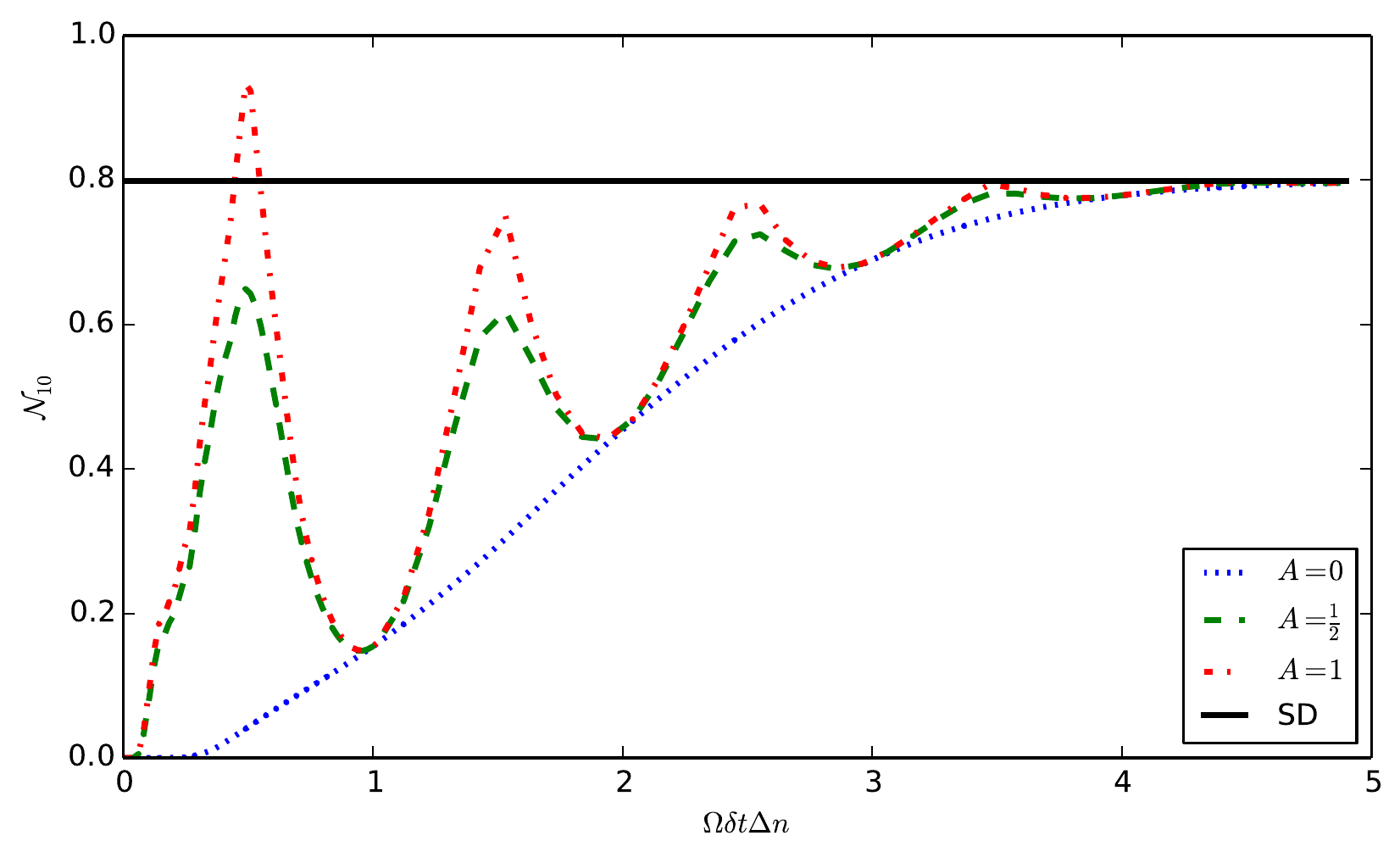}
  \caption{\label{fig:QWNM_10} Non-Markovianity of the Quantum walk after ten time steps, $\mathcal{N}_{10}$.
  Parameter $A$ controls the structure of the environment spectrum. See text for other parameter values.
  Non-Markovianity is plotted as a
  function of the dimensionless interaction time of each dephasing unit. As $\delta t\Omega\Delta n$ increases the 
  dephasing effect of each $U_{\delta t}$ becomes stronger and the solid black line corresponds to the strong 
  dephasing limit.}
\end{figure}

\subsubsection{Non-Markovianity of the open quantum walk}\label{sec:non-mark-open}

We choose the initial state pair for the walk to be $\ket{L}\ket{0}$ and $\ket{R}\ket{0}$ for simplicity.
Parameters of the environment are $\delta\omega = 9\sigma$, $\Delta n = 0.009$, and we vary the value of parameter
$A$ which controls the structure of the environment. 
 Non-Markovianity as 
a function of the dimensionless parameter $\delta t \Omega \Delta n$ is  plotted in Fig.~\ref{fig:QWNM_10},
where $\Omega = \frac{\delta\omega}{2\pi}$. 

We can see that the measure for 
non-Markovianity is periodic when we increase $\delta t$ which gives the duration of the dephasing step.
Remarkably, for a specific values of $\delta t$ the structure of the environment is irrelevant, i.e., for specific duration of dephasing,
the quantity of memory effects remain the same irrespective of the structure of the environment. 
The periodicity of the 
measure is related to the separation of the peaks of the environment spectrum in the sense that 
$\mathcal{N}\vert_{A=0}=\mathcal{N}\vert_{A=a}$ when $\delta t = m \frac{2\pi}{\delta\omega \Delta n}$.
It is also very interesting to note that for strong dephasing limit, the value of non-Markovianity approaches a constant value which is again independent of the structure of the environment which resembles the simple discrete qubit dynamics case. 
However, before the strong dephasing limit when $\delta t$ has small or intermediate value, non-Markovianity has two sources: the local unitary and the environmental structure.
For $A=0$, the non-Markovianity originates from the coin flipping only (local unitary) while for other presented values of  $A$, the structure of the environment gives additional contribution to the quantity of the memory effects.
 
\section{Summary}\label{sec:summary-1}
We have analyzed in detail two different models for discrete open quantum system dynamics: a simple qubit dynamics with dephasing and more elaborate full open quantum walk model. The results show that
the addition of local unitary operation can dramatically change the nature of the open system dynamics and in particular the appearance of memory effects. 
We have discussed how memory effects generated within our models can become generic in the sense that 
the structure of the environment spectrum does not play a role -- the amount of memory effects is independent of the form of the spectra -- and non-Markovianity is solely induced by local unitary operation in these cases. The considered models and observed phenomena give rise to 
approximation schemes which simplify the description of the dynamics considerably that may prove useful for 
other contexts  too. We have also discussed in detail for the first time a model for open quantum walk where memory effects can be introduced
and controlled in an experimentally relevant way. 
The presented open quantum walk is experimentally realizable using linear optics. 

\begin{acknowledgements}
The authors thank Vilho, Yrj\"o, and Kalle V\"ais\"al\"a 
Foundation, and Jenny and Antti Wihuri Foundation for financial support 
and Chuan-Feng Li for useful discussions.
This research was undertaken on Finnish Grid Infrastructure (FGI) resources.
\end{acknowledgements}

\bibliography{QW}
\end{document}